\documentclass{aastex}          
\usepackage{spr-astr-addons}    
\usepackage{natbib}             

\newcommand{\tens}{\mathbf}
\begin{document}
\newlength{\tmpl}
\newlength{\tmplj}
\newlength{\tmpla}
\newlength{\tmplb}
\newlength{\tmplc}
\newlength{\tmpld}
\newlength{\tmple}
\newlength{\tmplf}
\newlength{\tmplg}
\newlength{\tmplh}
\newlength{\tmpli}

\title{Large-scale collective motion of RFGC galaxies}

\shorttitle{Large-scale collective motion of RFGC galaxies}
\shortauthors{S. L. Parnovsky \and A. S. Parnowski}

\author{S. L. Parnovsky} 
\affil{Astronomical Observatory of Taras Shevchenko Kyiv National University\\
prov. Observatorny, 3, 04053, Kyiv, Ukraine\\
tel: +380444860021, fax: +380444862191\\ e-mail:par@observ.univ.kiev.ua}
\email{par@observ.univ.kiev.ua}

\author{A. S. Parnowski} 
\affil{Space Research Institute of NASU and NSAU\\
prosp. Akad. Glushkova, 40, korp. 4/1, 03680 MSP, Kyiv-187, Ukraine\\
tel: +380933264229, fax: +380445264124\\ e-mail:parnowski@gmail.com}
\email{parnowski@gmail.com}

\begin{abstract}
We processed the data about radial velocities and HI linewidths for $1678$ flat
edge-on spirals from the Revised Flat Galaxy Catalogue. We obtained the
parameters of the multipole components of large-scale velocity field of
collective non-Hubble galaxy motion as well as the parameters of the
generalized Tully-Fisher relationship in the ``HI line width -- linear
diameter'' version. All the calculations were performed
independently in the framework of three models, where the multipole decomposition
of the galaxy velocity field was limited to a dipole, quadrupole and octopole
terms respectively. We showed that both the quadrupole and the octopole
components are statistically significant.

On the basis of the compiled list of peculiar velocities of $1623$ galaxies we
obtained the estimations of cosmological parameters $\Omega_m$ and $\sigma_8$.
This estimation is obtained in both graphical form and as a constraint of the
value $S_8=(\Omega_m/0.3)^{0.35} \sigma_8 = 0.91 \pm 0.05$.
\end{abstract}


\section{Introduction}\label{s:Introduction}

The distribution of matter, including dark matter, in the Universe is inhomogeneous
on the scales of about $100\,Mpc$. This is manifested, for example, in the
existence of superclusters and voids. A galaxy, besides the cosmological expansion,
is also attracted to the regions with greater density. As a result, the galaxies are
involved in a non-Hubble large-scale collective motion on the background of
Hubble expansion. In a more sophisticated way this can be considered as a
development of initial density and velocity fluctuations in the early Universe
due to the gravitational instability.

Investigation of such a motion is important
at least for two reasons. First of all, it allows to plot the distribution of
matter, including dark matter, in the surrounding region of the Universe, which
is the main goal of cosmography, and to compare this distribution with the
distribution of luminous matter. The second reason is that the parameters of
this motion are linked with certain cosmological parameters, so we can obtain
new independent estimations of these parameters. Of course, the accuracy of such
estimation will not be very high, but the agreement of different estimations
of cosmological parameters can support the correctness of the standard cosmological
model.

In 1989 \citeauthor{ref:K89} proposed to use flat edge-on spiral galaxies as a tool
for studying their large-scale collective motion. They are good in this role for the
following reasons:
\begin{enumerate}
\item{The linear diameter is strongly correlated with the HI linewidth for thin
bulgeless galaxies. This allows to determine distances without photometric data.}
\item{Flat galaxies can be easily identified by their axes ratio.}
\item{Flat galaxies have a nearly 100 per cent HI detection rate.}
\item{Galaxies in clusters are usually not flat due to interaction with neighbours.
This means that flat galaxies avoid clustering and do not interact with the
intergalactic gas in clusters.}
\item{Peculiar velocities of isolated flat galaxies are not perturbed by neighbours.}
\end{enumerate}

To implement this idea the Flat
Galaxy Catalogue \citep[FGC,][]{ref:FGC} was created. It contained data about $N=4455$
galaxies, which satisfied the conditions $a_b/b_b\ge7$ and $a_b>0.6'$. Here
$a_b$ and $b_b$ are the major and minor axial diameters directly measured on
POSS-I and ESO/SERC plates. In accordance with the original photographic
material, the Catalogue consists of two parts: FGC ($N=2573$) and its southern
extension, FGCE ($N=1882$). The first part is based on the POSS-I and covers the
sky region with declinations between $-20\deg$ and $+90\deg$. The second one is
based on the ESO/SERC and covers the rest of the sky area up to $\delta=-90\deg$.

After thorough studies of the catalogue's properties, both
these parts were joined. The angular diameters from the FGCE were converted to the
POSS-I system of the FGC, which appeared to be close to the $a_{25}$ system.
This system, where galaxy size is taken at $B=25\,mag/\square''$ isophotal
level, was used, in particular, by \citet{ref:deVac}. Some FGCE galaxies, which
did not satisfy the condition $a>0.6'$ after conversion, were removed from the
resulting Revised Flat Galaxy Catalogue \citep[RFGC,][]{ref:RFGC}. It contained data
about $N=4236$ galaxies including the information on the following parameters:
Right Ascension and Declination for the epochs J2000.0 and B1950.0, galactic longitude
and latitude, major and minor blue and red diameters in arcminutes in the POSS-I diameter
system, morphological type of the spiral galaxies according to the Hubble
classification, index of the mean surface brightness (I -- high, IV -- very low)
and some other parameters, which are not used in this article. More detailed
description of the catalogue can be found in the paper \citep{ref:RFGC}.

The original goal of this catalogue was to estimate the distance to galaxies
according to the Tully-Fisher relationship in the ``HI line width -- linear
diameter'' version without using their redshifts. The difference between the
velocity $V$ derived from the redshift and the Hubble velocity $R=Hr$
corresponding to the distance $r$ estimated by Tully-Fisher relationship is
called a peculiar velocity $V_{pec}=cz-Hr$. We can use such a simple form of the
Hubble's law because our sample has $z<0.1$.

There are some important things to take
into account about peculiar velocities. The redshift includes not only the
velocity of the galaxy, but also the velocities of our Galaxy, Solar System and
the Earth. Thus, to eliminate these factors, all velocities were reduced to the
frame, where CMB is isotropic. Naturally, the redshift gives us only the radial
component of the velocity and the tangential components cannot be measured.
Additionally, Tully-Fisher relationship is statistical and thus has a certain
error. Thus, we can only estimate the peculiar velocity for each galaxy,
sometimes with a significant error.

However, we believe that there is a large-scale velocity
field. We consider the individual galaxies as test particles in the velocity
field of large-scale collective motion. Thus, having data about the peculiar
velocities of a large number of galaxies, we can restore the radial component of
the velocity field. Using some additional assumptions, like the potentiality of
the flux, it is possible to restore the 3D velocity field. For this reason we
need ample samples of peculiar velocities, preferably uniformly covering the
celestial sphere.

\begin{figure*}[tb]
\includegraphics[width=\textwidth]{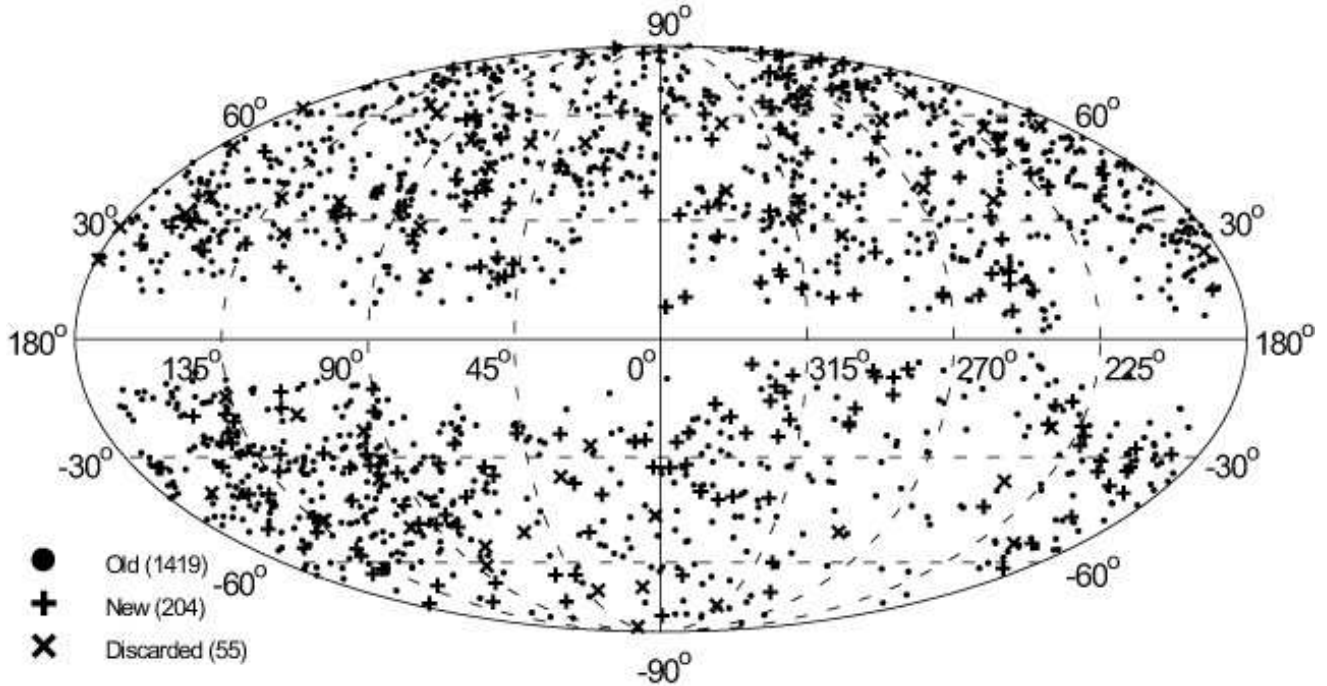}
\caption{Distribution of galaxies over the celestial sphere in galactic coordinates}
\label{fig:1}
\end{figure*}

These conditions are satisfied by the RFGC catalogue, which covers the whole
celestial sphere in both hemispheres with the natural exclusion of the Milky Way
region. However, not all of the RFGC galaxies have data required to estimate the
distance to them using the Tully-Fisher relationship. Nevertheless, the sample of
galaxies having such data is also quite uniform, as shown on Fig.
\ref{fig:1}.

Naturally, RFGC is not the only sample that can be used for
this purpose. Possible alternatives include SBF \citep{ref:SBF97,ref:SBF00},
ENEAR \citep{ref:ENEAR}, SFI \citep{ref:SFIa,ref:SFIb}, SFI++ \citep{ref:SFIpp},
EFAR \citep{ref:EFAR96,ref:EFAR01}, SMAC \citep{ref:SMAC99,ref:SMAC04},
2MFGC \citep{ref:2MFGC}. However, in this article we use only the RFGC catalogue.

To apply the Tully-Fisher relationship the data from the catalogue is not sufficient;
we also need to know the HI linewidth $W$ (in this article we take it at $50$ per cent
level), or the gas rotation velocity $V_{rot}$ obtained from optical
observations. Additionally, we need redshift data. The number of galaxies with
such data increases constantly.

Since FGC and RFGC were assembled, a number of articles were published dealing with
collective motions of RFGC galaxies. Very preliminary results were published by
\citet{ref:K95}. The parameters of the bulk motion were calculated by
\citet{ref:K00}. In the paper \\ \citep{ref:Par01} not only new data were added, but
also the models featuring the quadrupole and octopole components of the
velocity field were proposed. Also, the generalized Tully-Fisher relationship
for RFGC galaxies was finalized. It included data not only about
HI linewidth and angular diameter in red and blue bands, but also about the
morphological type of the galaxy and its surface brightness index. By that time
the authors had information about radial velocities and HI $21\,cm$ line widths
or $V_{rot}$ for $1327$ RFGC galaxies from the different
sources listed in the paper \citep{ref:K00}. From this number, $1271$ galaxies were
included in the sample, and the rest of them were considered to be outliers. As
a result, the first list of peculiar velocities of RFGC galaxies was prepared by
\citet{ref:list}. Four years later, the number of galaxies with available data
increased and reached $1561$ \citep{ref:ParTug04}; $1493$ of them entered the
sample. A new version
of the list of peculiar velocities was prepared by \citet{ref:ParTug05}. This
list was the basis for solving the two abovementioned problems, namely mapping the
matter density and estimation of cosmological parameters. The distribution
of matter density was obtained in the paper \citep{ref:SharPar} up to
$75h^{-1}\,Mpc$ in the supergalactical plane and 8 more planes. In the same
article, the excess masses of attractors in this region were estimated, and the
estimation of $\beta$ parameter was obtained. The estimations of the cosmological
parameters $\Omega_m$ and $\sigma_8$ were obtained in the paper \citep{ref:cosm}.
The model was expanded taking into account the effects of general theory of relativity
in the paper \citep{ref:ParGayd05}.

An important result, obtained in the paper \\ \citep{ref:Par01}, and confirmed by
\\ \citet{ref:ParTug04}, was the statistical significance of quadrupole and octopole
components (both at more than $99$ per cent confidence level). This result relied
on the F-test, which assumes normal distribution of errors \citep[see][]{ref:Hudson}.
However, even the normally distributed errors of angular diameters and HI
linewidths and deviations from Tully-Fisher relationship lead to non-Gaussian
distribution of peculiar velocities. Thus, the statistical significance of these
terms required additional consideration. In the article \citep{ref:ParPar08} it
was shown using Monte-Carlo simulations that the statistical significance of
these components appeared to be less than perceived from the F-test, but still large
enough to be considered: the quadrupole was statistically significant at $96$
per cent confidence level and the octopole -- at $90..93$ per cent confidence
level.

In the four years that passed since the last list of peculiar velocities of RFGC
galaxies was compiled, not only new data have appeared, but also some old data
were remeasured. This led to a necessity of repeating all the steps required to
obtain the new list. This means that not only new data have to be added, but the
process of sample selection must be repeated from the very beginning. As it will
be shown further, some results appeared to be notably different from the
ones obtained with the previous sample.

In this paper we present the parameters of the large-scale collective velocity
field in the framework of the multipole model as well as the estimations of some
cosmological parameters. The distribution of matter density will be discussed
in a separate article.

\section{Multipole models of collective motion of galaxies}\label{s:Models}

In the paper \citep{ref:Par01} the velocity field components in the CMB reference
frame were expanded in terms of galaxy's radial vector $\vec{r}$. The first
three terms are
\begin{equation}\label{eqn:1}
V_{i}=D_{i}+A_{ik}r_{k}+O'_{ikl}r_{k}r_{l}.
\end{equation}
In our notation we use the Einstein rule: summation by all the repeating
indices. Hence, for the radial velocity $V_{3K}$ in the CMB reference
frame we get
\begin{equation}\label{eqn:2}
V_{3K}=D_{i}n_{i}+rA_{ik}n_{i}n_{k}+r^{2}O'_{ikl}n_{i}n_{k}n_{l},\,
\end{equation}
where $\vec{n}=\vec{r}/r$ and $r=|\vec{r}|$. Let us decompose the tensor
$\tens{A}$ in two parts: $A_{ik}=H\delta_{ik}+Q'_{ik}$. Here $H$ is a trace,
corresponding to the Hubble constant, $\delta_{ik}$ is the Kronecker delta, and
$\tens{Q}'$ is a traceless tensor. Next we switch from distance $r$ to the
corresponding Hubble velocity $R=Hr$.
\begin{equation}\label{eqn:3}
V_{3K}=R+D_{i}n_{i}+RQ_{ik}n_{i}n_{k}+R^{2}O_{ikl}n_{i}n_{k}n_{l},
\end{equation}
where $\tens{Q}=H^{-1}\tens{Q}'$ and $\tens{O}=H^{-2}\tens{O}'$.

This decomposition was used to obtain some models of dependence of galaxy's
radial velocity $V$. In the simplest D-model (Hubble law + dipole) we have
\begin{equation}\label{eqn:D}
V_{3K}=R+V^{dip}+\delta V,\,V^{dip}=D_{i}n_{i},
\end{equation}
where $\vec{D}$ is a velocity of homogeneous bulk motion, $\delta V$ is a
random deviation and $\vec{n}$ is a unit vector towards galaxy. After the
addition of quadrupole terms we obtain a DQ-model
\begin{equation}\label{eqn:DQ}
V_{3K}=R+V^{dip}+V^{qua}+\delta V,\,V^{qua}=RQ_{ik}n_{i}n_{k}
\end{equation}
with symmetrical traceless tensor $\tens{Q}$ describing quadrupole components of
the velocity field. It characterises the relative deviation of an effective
``Hubble constant'' in a given direction from the mean value. More detailed
discussion will be given further. The DQO-model includes the octopole component
of velocity field described by a symmetrical tensor $\tens{O}$ of rank 3:
\begin{equation}\label{eqn:DQO}
\begin{array}{l}
V_{3K}=R+V^{dip}+V^{qua}+V^{oct}+\delta V,\\
V^{oct}=R^{2}O_{ikl}n_{i}n_{k}n_{l}.
\end{array}
\end{equation}
In some cases it makes sense to use another way of describing the octopole
component in the DQO-model. For this purpose the tensor $\tens{O}$ can be
considered a sum of a traceless tensor $\tens{\hat{O}}$, and a trace,
characterized by the vector $\vec{P}$
\begin{equation}\label{eqn:P}
O_{ijk}=\hat{O}_{ijk}+P_{(i}\delta_{jk)}, P_i=\frac{3}{5}O_{ijk}\delta_{jk}.
\end{equation}
Here the indices in parantheses are symmetrized. Thus an alternative form of the
DQO-model is
\begin{equation}\label{eqn:DQOP}
\begin{array}{l}
V_{3K}=R+ (D_{i}+R^{2}P_{i})n_{i}+ RQ_{ik}n_{i}n_{k}\\
\phantom{V_{3K}=}{}+R^{2}\hat{O}_{ikl}n_{i}n_{k}n_{l}+\delta V.
\end{array}
\end{equation}
We will use Cartesian components of the vector $\vec{n}$ in the galactic
coordinates: 
\begin{equation}\label{eqn:lb}
\begin{array}{l}
n_1=n_z=\sin b, n_2=n_x=\cos l \cos b,\\
n_3=n_y=\sin l \cos b.
\end{array}
\end{equation}

These three models were used for processing data on RFGC galaxies. To estimate
the distances to galaxies a generalized Tully-Fisher relationship was used in
the `linear diameter -- HI line width' version. It has a form \citep{ref:Par01}
\begin{equation}\label{eqn:TF}
\begin{array}{l}
R=(C_1+C_2B+C_3BT+C_4U)\frac{W}{a}\\
\phantom{R=}{}+C_5\left(\frac{W}{a}\right)^2+C_6\frac{1}{a},
\end{array}
\end{equation}
where $W$ is a corrected HI line width in $km\,s^{-1}$ measured at $50$ per cent
of the maximum, $a$ is a corrected major galaxies' angular diameter in arcminutes
on red POSS and ESO/SERC reproductions, $U$ is a ratio of major galaxies'
angular diameters on red and blue reproductions, $T$ is a morphological type
indicator ($T=I_{t}-5.35$, where $I_{t}$ is a Hubble type; $I_{t}=5$ corresponds
to type Sc), and $B$ is a surface brightness indicator ($B=I_{SB}-2$, where
$I_{SB}$ is a surface brightness index from RFGC; brightness decreases from I to
IV). Details of corrections one can find in the papers \citep{ref:Par01,ref:ParTug04}.
The reasons for choosing this form of Tully-Fisher
relationship for RFGC galaxies are given ibid. We only note that the statistical
significance of each term in eq. (\ref{eqn:TF}) is greater than $99$ per cent.

Note that eq. (\ref{eqn:TF}) differs from the classical Tully-Fisher relationship,
which has the form
\begin{equation}\label{eqn:TFalpha}
R=C\frac{W^\alpha}{a}.
\end{equation}
The corresponding analysis was performed in the paper \citep{ref:ParGayd05} for
the relativistic model of motion. It was shown that the power $\alpha$ for
the classical form of Tully-Fisher relationship (\ref{eqn:TFalpha}) somewhat
differs from 1. However, after including the corrections for the surface
brightness and the morphological type of the galaxy, the value of $\alpha$
appeared to be close enough to 1 to use the form given by (\ref{eqn:TF}). If we
repeat this analysis for the new data, the minimal variance is reached at
$\alpha\sim 1.16$. At the same time the standard deviation $\sigma$ for the form
(\ref{eqn:TFalpha}) is
$5..7$ per cent more than for the form (\ref{eqn:TF}) depending on the depth of
the subsample and model used. If we combine the forms (\ref{eqn:TF}) and
(\ref{eqn:TFalpha}) into
\begin{equation}\label{eqn:TFfull}
\begin{array}{l}
R=(C_1+C_2B+C_3BT+C_4U)\frac{W^\alpha}{a}\\
\phantom{R=}{}+C_5\left(\frac{W^\alpha}{a}\right)^2+C_6\frac{1}{a},
\end{array}
\end{equation}
we obtain the minimal variance at $\alpha\sim 1.05$, which is very close to 1.
This correction does not reduce $\sigma$ due to the decreased number of degrees
of freedom. This makes the introduction of $\alpha$ senseless. For this reason
we will use only the form (\ref{eqn:TF}). This is additionally justified by the
fact that all the coefficients enter this equation linearly and thus we can use
the linear regression analysis.

The D-model has 9 parameters (6 coefficients $C$ and 3 components of vector
$\vec{D}$), DQ-model has 14 parameters (5 components of tensor $\tens{Q}$ are
added), and DQO-model is described by 24 coefficients. The values and errors
of the coefficients were calculated by the least square method for different
subsamples with distances limitation to make the sample more homogeneous in
the depth \citep{ref:Par01}. We used the same weights for all datapoints. Since
the quadrupole and octopole terms explicitly depend on $R$, an iteration procedure
was used. Note that the coefficients of the generalised Tully-Fisher relationship
and components of the velocity model were fitted simultaneously. The iterations
converge rather quickly.

\section{Description of data}\label{s:Data}

Directly from the RFGC catalogue we take the angular diameters on blue and red
reproductions, morphological type of a galaxy and its surface brightness index.
The angular diameters are corrected for intrinsic absorption. From other sources
we take the radial velocity in the CMB frame and the HI line width or the gas
rotation velocity $V_{rot}$ obtained from optical observations. The line widths are
corrected for cosmological expansion and for turbulence. The data on HI line
widths for RFGC galaxies in the HyperLeda \citep{ref:LEDA} catalogue are based on
149 sources, which have different reliability. The main sources used
in previous versions of our samples are listed in the papers \citep{ref:Par01,ref:ParTug04,
ref:list,ref:K00,ref:ParTug05}. Since that time new large volumes of data on radial
velocities and HI linewidths of galaxies were published \citep[e.g.,][]{ref:Springob}.
For this reason we rechecked the
original data on HI line widths of galaxies in previous samples and changed some
of them. For each galaxy we used only the data of original measurements and did
not average the data of different papers. In the cases when data from reliable
sources differed significantly, we used less reliable sources to choose between
them. We also checked the deviation of the distance to galaxy from the
Tully-Fisher relationship in the D-model.

After revising data from the previous samples, we added new data measured during
the last 5 years. We took HI linewidths from the original articles
\citep{ref:31011,ref:31012,ref:31020,ref:51007}. In the case
when $W$ was unavailable we converted $V_{rot}$ from the HyperLeda catalogue
\citep{ref:Vrot} to $W$ using the relationship $W = 2.016 * V_{rot} + 20.6$,
obtained by processing data for galaxies, which have both records. Note that
there are rare cases of significant deviations from this relationship.

In the process of data preparation we constantly checked the deviations of the
radial velocities from the velocities calculated in the D-model. Some data were
obvious outliers. Such galaxies were rejected from the sample. As a result, we
had all the necessary data for $1678$ galaxies, including $55$ outliers. Thus the
sample contained $1623$ galaxies, including $204$ new and $1419$ old ones. It is noteworthy
that some of the old data changed as well. The distribution of these galaxies
over the celestial sphere is
shown on Fig. \ref{fig:1}. The mean radial velocity of old galaxies is
$6000\,km\,s^{-1}$, and of new ones it is $5600\,km\,s^{-1}$. The mode in both
cases is about $4800\,km\,s^{-1}$. The sample has $90\%$ completeness up to
about $6000\,km\,s^{-1}$.

The farthest galaxies in the sample have radial velocities above
$20000\,km\,s^{-1}$. However, at such scales the sample is inhomogeneous and
incomplete. To deal with more homogeneous samples, we consider subsamples with
distance limitation. They include all the galaxies satisfying the condition
$R<R_{max}$ in the D-model and are used for all the three models. For this purpose
we used the following iterative procedure.  First, we calculate the distances to the
galaxies using seed coefficients from the whole sample or another subsample with
close $R_{max}$. Then we select the galaxies with $R<R_{max}$ and recalculate the
coefficients using only their data, once again calculate the distances and so on.
In most cases these iterations converge, but in some cases, more often for small
$R_{max}$, we get a limit cycle. In this case we select only the galaxies whose
distances are below $R_{max}$ for all sets of coefficients in this limit cycle.

We used different subsamples with $R_{max}$ ranging from $3000$ to
$12000\,km\,s^{-1}$. In this paper we mostly present information for the
subsamples with $R_{max}$ equal to $8000\,km\,s^{-1}$ ($r=80h^{-1}\,Mpc$) and
$10000\,km\,s^{-1}$ ($r=100h^{-1}\,Mpc$).

\section{Multipole structure of collective motion}\label{s:Results}

The parameters of a generalized Tully-Fisher relationship for all three models and
two $R_{max}$ values are given in Table \ref{tbl:1} together with their statistical
significance levels according to F-test. The $F$ values should be compared to
$3.8$, $6.6$, $7.9$, $10.8$ and $12.1$, which correspond to $95$, $99$, $99.5$,
$99.9$ and $99.95$ per cent confidence levels respectively. Most of them are
statistically significant at $99.95$ per cent confidence level with the exception
of $C_3$ for $R_{max}=10000\,km\,s^{-1}$. Comparing the coefficients to that of
\citet{ref:ParTug04} one can see that $C_1$, $C_4$, $C_5$ and $C_6$ remain the same
within the margin of error, and $C_2$ and especially $C_3$ are reduced, which
means lesser influence of the morphological type of galaxies. This also leads to
some decrease of statistical significance, which is still more than $95$ per
cent.

\begin{table*}[tb]
\settowidth{\tmpl}{$6$}
\settowidth{\tmplj}{$\pm$}
\settowidth{\tmpla}{$(-5.54$}
\settowidth{\tmplb}{$1.53)\times 10^{-3}$}
\settowidth{\tmplc}{$92.4$}
\settowidth{\tmpld}{$(-6.63$}
\settowidth{\tmple}{$1.53)\times 10^{-3}$}
\settowidth{\tmplf}{$90.0$}
\settowidth{\tmplg}{$(-7.85$}
\settowidth{\tmplh}{$1.57)\times 10^{-3}$}
\settowidth{\tmpli}{$88.4$}
\caption{Parameters of the generalized Tully-Fisher relationship (\ref{eqn:TF})}
\label{tbl:1}
\begin{tabular}{r@{}r@{~$\pm$~}l|c|r@{~$\pm$~}l|c|r@{~$\pm$~}l|c}
\tableline
\hbox to \tmpl {$i$}&
\multicolumn{9}{c@{}}{\begin{tabular}{|r@{~\hspace*{\tmplj}~}l|c|r@{~\hspace*{\tmplj}~}l|c|r@{~\hspace*{\tmplj}~}l|c}
\multicolumn{3}{|c|}{D-model}&\multicolumn{3}{c|}{DQ-model}&\multicolumn{3}{c}{DQO-model}\\
\tableline
\hbox to \tmpla {}&\hbox to \tmplb {$C_i$}&\hbox to \tmplc {$F_i$}&\hbox to \tmpld {}&\hbox to \tmple {$c_i$}&\hbox to \tmplf {$F_i$}&\hbox to \tmplg {}&\hbox to \tmplh {$c_i$}&\hbox to \tmpli {$F_i$}\\
\end{tabular}}\\
\tableline
\end{tabular}
\begin{tabular}{r|r@{~$\pm$~}l|c|r@{~$\pm$~}l|c|r@{~$\pm$~}l|c}
\multicolumn{10}{c}{$R_{max}=8000$ km/s (1240 galaxies)}\\
$1$&$15.8$&$1.6$&$103$&$17.2$&$1.6$&$122$&$17.4$&$1.6$&$126$\\
$2$&$1.84$&$0.20$&$80.5$&$1.96$&$0.21$&$90.0$&$1.95$&$0.21$&$88.4$\\
$3$&$-0.46$&$0.12$&$13.4$&$-0.49$&$0.13$&$15.2$&$-0.45$&$0.13$&$12.9$\\
$4$&$8.29$&$1.33$&$38.6$&$7.13$&$1.34$&$28.6$&$7.24$&$1.33$&$29.6$\\
$5$&$(-5.54$&$1.53)\times 10^{-3}$&$13.1$&$(-6.63$&$1.53)\times 10^{-3}$&$18.7$&$(-7.85$&$1.57)\times 10^{-3}$&$25.0$\\
$6$&$-825$&$86$&$92.4$&$-867$&$85$&$103$&$-883$&$85$&$107$\\
\multicolumn{10}{c}{$R_{max}=10000$ km/s (1459 galaxies)}\\
$1$&$16.3$&$1.4$&$143$&$17.3$&$1.4$&$157$&$17.2$&$1.4$&$156$\\
$2$&$1.38$&$0.18$&$59.3$&$1.44$&$0.18$&$62.1$&$1.55$&$0.18$&$70.6$\\
$3$&$-0.24$&$0.11$&$4.55$&$-0.26$&$0.11$&$5.32$&$-0.27$&$0.11$&$5.54$\\
$4$&$7.62$&$1.17$&$42.4$&$6.80$&$1.18$&$33.0$&$6.99$&$1.18$&$35.3$\\
$5$&$(-5.96$&$1.05)\times 10^{-3}$&$32.0$&$(-6.41$&$1.06)\times 10^{-3}$&$36.4$&$(-6.71$&$1.13)\times 10^{-3}$&$34.9$\\
$6$&$-785$&$87$&$81.8$&$-817$&$87$&$88.2$&$-845$&$87$&$94.9$\\
\tableline
\end{tabular}
\end{table*}

\begin{figure}[tb]
\includegraphics[width=\columnwidth]{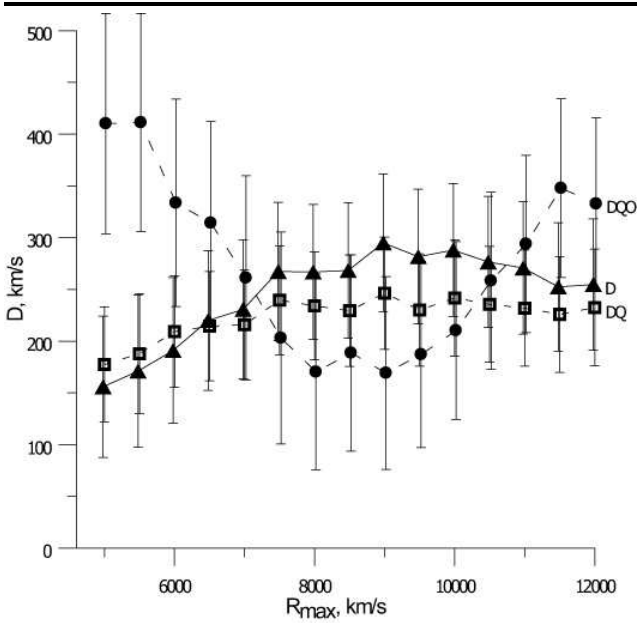}
\caption{Dependence of the dipole component of the velocity on the sample depth
for 3 models. Triangles correspond to the D-model, squares -- to the DQ-model,
and circles -- to the DQO-model}
\label{fig:2}
\end{figure}

\begin{figure}[tb]
\includegraphics[width=\columnwidth]{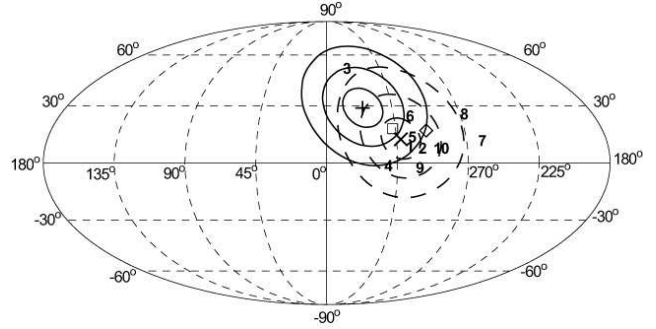}
\caption{Bulk motion apices in galactic coordinates for $R_{max}=10000\,km\,s^{-1}$.
Crosses mark the apices of the bulk motion in the D-model surrounded by $1\sigma$,
$2\sigma$ and $3\sigma$ confidence areas. The square marks the apex position of
the galaxies from the previous sample with corrected data. The diamond marks the
apex position in the DQO-model. Solid boundaries correspond to new results,
and dashed ones -- to the results of \citet{ref:ParTug04}.
Numbers denote the results of other authors: 1 -- \citep{ref:LB}, 2 --
\citep{ref:H95}, 3 -- \citep{ref:LP}, 4 -- \citep{ref:Par01}, 5 --
\citep{ref:Dekel99}, 6 -- \citep{ref:daCosta00}, 7 -- \citep{ref:SMAC04}, 8 -- 
\citep{ref:Dale99}, 9 -- \citep{ref:Kudrya03}, 10 -- \citep{ref:WFH09}}
\label{fig:3}
\end{figure}

The inclusion of the octopole component for subsamples with $R_{max}$ about
$8000\,km\,s^{-1}$ leads to a drastic decrease of the dipole component up to
$1\sigma$ level, as \citet{ref:Par01} have shown. This effect appeared to be even
more evident in the paper \citep{ref:ParTug04}. Such a strong dipole-octopole
coupling can be explained by the same symmetry of these components with respect
to the inversion of space. For the new sample this effect appeared to be weaker as
the dipole component decreases only up to $2\sigma$ level (see Fig. \ref{fig:2}).
Naturally this effect is caused by incompleteness and asymmetry of the sample,
which lead to dipole and octopole components being non-orthogonal.

This effect is important because D and DQO models yield the values of the bulk 
motion velocity differing by a factor of 2. Thus the dependence of the bulk motion
velocity on the sample depth will be totally different for D and DQO models as
one can see from Fig. \ref{fig:2}. One should take this into account when
analysing the convergence scale.

In comparison with the papers \citep{ref:Par01} and \citep{ref:ParTug04}, the apex
direction of the dipole component has changed. In Table \ref{tbl:2} we present its
parameters for 2 subsamples together with the standard deviation $\sigma$ for D-model.
Note that $\sigma$ includes both the intrinsic scatter in the generalised Tully-Fisher
relationship and the stochastic component of the velocities, resulting in high values.

\begin{table*}[tb]
\settowidth{\tmpl}{10000}
\settowidth{\tmplj}{$\pm$}
\settowidth{\tmpla}{$-59$ $\pm$ $79$}
\settowidth{\tmplb}{$l$, deg}
\settowidth{\tmplc}{$229$ $\pm$ $104$}
\settowidth{\tmpld}{$b$, deg}
\settowidth{\tmple}{$-182$ $\pm$ $100$}
\settowidth{\tmplf}{$\sigma$, km/s}
\settowidth{\tmplg}{$288$ $\pm$ $95$}
\settowidth{\tmpli}{Model}
\caption{Parameters of the dipole component}
\label{tbl:2}
\begin{tabular}{l@{}r|r@{~$\pm$~}l|r@{~$\pm$~}l|r@{~$\pm$~}l|r@{~$\pm$~}l@{}r|r|r}
\tableline
Model&\multicolumn{9}{c}{\begin{tabular}{|c|c|c|c|c|}
\hbox to \tmpl {\hfil $R_{max}\hfil$}&\hbox to \tmpla {\hfil $D_z$\hfil}&\hbox to \tmplc {\hfil $D_x$\hfil}&\hbox to \tmple {\hfil $D_y$\hfil}&\hbox to \tmplg {\hfil $D$\hfil}\\
\tableline
\multicolumn{5}{|c|}{km/s}
\end{tabular}}&$l$, deg&$b$, deg&$\sigma$, km/s\\
\tableline
\end{tabular}
\begin{tabular}{l|r|r@{~$\pm$~}l|r@{~$\pm$~}l|r@{~$\pm$~}l|r@{~$\pm$~}l|r|r|r}
D&$8000$&$147$&$47$&$176$&$58$&$-136$&$56$&$267$&$65$&$322$&$33$&$1045$\\
DQ&$8000$&$81$&$50$&$174$&$60$&$-134$&$58$&$234$&$52$&$323$&$20$&$1033$\\
\hbox to \tmpli {DQO}&$8000$&$-59$&$79$&$122$&$104$&$-103$&$100$&$170$&$95$&\hbox to \tmplb {\hfil $320$}&\hbox to \tmpld {\hfil $-20$}&\hbox to \tmplf {\hfil $1020$}\\
D&$10000$&$139$&$49$&$229$&$59$&$-106$&$59$&$288$&$64$&$335$&$29$&$1159$\\
DQ&$10000$&$70$&$53$&$200$&$62$&$-116$&$62$&$242$&$56$&$330$&$17$&$1151$\\
DQO&$10000$&$61$&$77$&$85$&$98$&$-182$&$97$&$210$&$86$&$295$&$17$&$1138$\\
\tableline
\end{tabular}
\end{table*}

\begin{figure}[tb]
\includegraphics[width=\columnwidth]{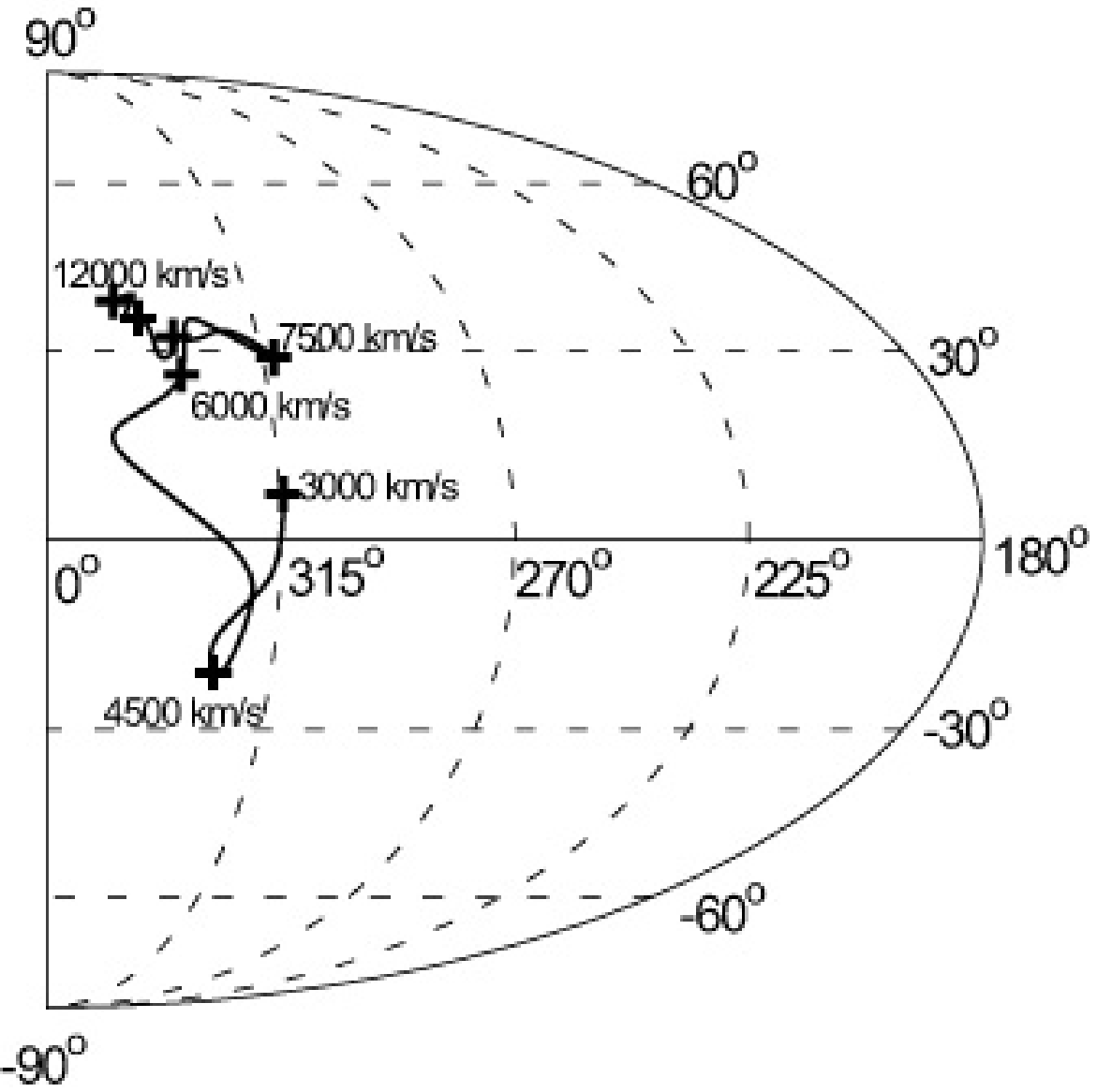}
\caption{Dependence of bulk motion apex on the sample depth}
\label{fig:4}
\end{figure}

The comparison of the results from Table \ref{fig:2} with those obtained earlier
\citep{ref:Par01,ref:ParTug04} shows that the norm of the velocity
hasn't changed significantly, but the apex direction has. Let us discuss this
question in detail. On Fig. \ref{fig:3} we present the Mollweide projection upon
the celestial sphere of the ellipsoids corresponding to $1\sigma$, $2\sigma$ and
$3\sigma$ confidence levels for the apex direction in the framework of D-model
for $R_{max}=10000\,km\,s^{-1}$. Solid boundaries correspond to new results, and
dashed ones -- to the results of \citet{ref:ParTug04}. One can see that the apex
direction has changed at slightly more than $2\sigma$. Besides, on the Fig
\ref{fig:3} we show the results of other authors: 1 -- \citep{ref:LB}, 2 --
\citep{ref:H95}, 3 -- \citep{ref:LP}, 4 -- \citep{ref:Par01}, 5 --
\citep{ref:Dekel99}, 6 -- \citep{ref:daCosta00}, 7 -- \citep{ref:SMAC04}, 8 -- 
\citep{ref:Dale99}, 9 -- \citep{ref:Kudrya03}, 10 -- \citep{ref:WFH09}. One can see that the
result of \citet{ref:ParTug04} matches other results better. The new result
became closer to the result of \citet{ref:LP}, but our norm is 2 times less.
What caused such a deviation? We found the apex for `old' galaxies used in
previous samples but with corrected data for the HI linewidth. At this stage the
apex has notably deviated from its original position. This position is shown on
Fig. \ref{fig:3} by a square. After adding new data this deviation increased.

Let us discuss how much this deviation affects the global picture of the large-scale
motion. Some authors believe that the norm and the apex of the dipole component are
the most prominent characteristics of large-scale motion. This was true for very
old samples with small depth. Our sample, however, has so large depth that it
includes most nearest superclasters like the Great Attractor and Perseus-Pisces.
For this reason, the first few components of the multipole decomposition
(\ref{eqn:3}) can not adequately describe all the radial velocity field of the
large-scale motion. Moreso, this is hard to do with a single component
(\ref{eqn:D}). This phenomenon is due to the fact that a multipole decomposition is capable of
describing very-large-scale motion with characteristic scale much larger than
the scale of the sample, but can not describe the not-so-large-scale motions
with characteristic scale less than the scale of the sample caused by
attraction to these superclusters for any reasonable number of components.

\begin{table*}[tb]
\settowidth{\tmpl}{Total}
\settowidth{\tmplj}{$\pm$}
\settowidth{\tmpla}{$-2.1$ $\pm$ $2.7$}
\settowidth{\tmplb}{$28.8$}
\settowidth{\tmplc}{$-1.9$ $\pm$ $2.8$}
\settowidth{\tmpld}{$22.6$}
\settowidth{\tmple}{$-1.9$ $\pm$ $2.5$}
\settowidth{\tmplf}{$4.63$}
\settowidth{\tmplg}{$-2.5$ $\pm$ $2.7$}
\settowidth{\tmplh}{$22.6$}
\caption{Coefficients of the quadrupole component}
\label{tbl:3}
\begin{tabular}{r@{}r@{~$\pm$~}l|r|r@{~$\pm$~}l|r|r@{~$\pm$~}l|r|r@{~$\pm$~}l|r}
\tableline
\hbox to \tmpl {\hfil $i$\hfil}&
\multicolumn{12}{c@{}}{\begin{tabular}{|c|c|c|c|c|c|c|c}
\multicolumn{4}{|c|}{$R_{max}=8000$ km/s}&\multicolumn{4}{c}{$R_{max}=10000$ km/s}\\
\tableline
\multicolumn{2}{|c|}{DQ-model}&\multicolumn{2}{c|}{DQO-model}&\multicolumn{2}{c|}{DQ-model}&\multicolumn{2}{c}{DQO-model}\\
\tableline
\hbox to \tmpla {\hfil $q_i,\,\%$\hfil}&\hbox to \tmplb {\hfil $F_i$\hfil}&\hbox to \tmplc {\hfil $q_i,\,\%$\hfil}&\hbox to \tmpld {\hfil $F_i$\hfil}&\hbox to \tmple {\hfil $q_i,\,\%$\hfil}&\hbox to \tmplf {\hfil $F_i$\hfil}&\hbox to \tmplg {\hfil $q_i,\,\%$\hfil}&\hbox to \tmplh {\hfil $F_i$\hfil}\\
\end{tabular}}\\
\tableline
\end{tabular}
\begin{tabular}{r|r@{~$\pm$~}l|r|r@{~$\pm$~}l|r|r@{~$\pm$~}l|r|r@{~$\pm$~}l|r}
1&$8.1$&$1.5$&$28.8$&$7.7$&$1.6$&$22.6$&$5.7$&$1.4$&$16.6$&$7.6$&$1.6$&$22.6$\\
2&$-2.1$&$1.6$&$1.6$&$-1.9$&$1.7$&$1.2$&$-1.9$&$1.5$&$1.7$&$-2.5$&$1.6$&$2.3$\\
3&$-0.8$&$2.2$&$0.1$&$-1.1$&$2.3$&$0.2$&$2.1$&$1.9$&$1.2$&$-1.2$&$2.1$&$0.3$\\
4&$0.9$&$2.3$&$0.1$&$1.2$&$2.5$&$0.2$&$4.1$&$2.1$&$4.0$&$2.6$&$2.5$&$1.1$\\
5&$1.7$&$2.7$&$0.4$&$0.1$&$2.8$&$0.0$&$1.7$&$2.5$&$0.5$&$-0.2$&$2.7$&$0.0$\\
Total&\multicolumn{2}{c|}{}&$6.44$&\multicolumn{2}{c|}{}&$5.10$&\multicolumn{2}{c|}{}&$4.63$&\multicolumn{2}{c|}{}&$4.85$\\
\tableline
\end{tabular}
\end{table*}

\begin{table*}[tb]
\settowidth{\tmpl}{$R_{max}$,}
\settowidth{\tmplj}{$1459$}
\settowidth{\tmpla}{$22.3$ $\pm$ $4.3$}
\settowidth{\tmplb}{$198$}
\settowidth{\tmplc}{$-9$}
\settowidth{\tmpld}{$-19.8$ $\pm$ $4.3$}
\settowidth{\tmple}{$109$}
\settowidth{\tmplf}{$-10$}
\settowidth{\tmplg}{$-3.0$ $\pm$ $2.7$}
\settowidth{\tmplh}{$359$}
\settowidth{\tmpli}{$-26$}
\caption{Eigenvalues and eigenvectors of the quadrupole tensor}
\label{tbl:4}
\begin{tabular}{r|r@{}r@{~$\pm$~}l|r|r|r@{~$\pm$~}l|r|r|r@{~$\pm$~}l|r|r}
\tableline
\begin{tabular}{@{}c@{}}$R_{max}$,\\km/s\\ \end{tabular}&\hbox to \tmplj {\hfil $N$\hfil}&
\multicolumn{12}{c@{}}{\begin{tabular}{|c|c|c|c|c|c|c|c|c}
\multicolumn{3}{|c|}{Maximum}&\multicolumn{3}{c|}{Minimum}&\multicolumn{3}{c}{Third axis}\\
\tableline
\hbox to \tmpla {\hfil $Q_1,\,\%$\hfil}&\hbox to \tmplb {\hfil $l$\hfil}&\hbox to \tmplc {\hfil $b$\hfil}&\hbox to \tmpld {\hfil $Q_2,\,\%$\hfil}&\hbox to \tmple {\hfil $l$\hfil}&\hbox to \tmplf {\hfil $b$\hfil}&\hbox to \tmplg {\hfil $Q_3,\,\%$\hfil}&\hbox to \tmplh {\hfil $l$\hfil}&\hbox to \tmpli {\hfil $b$\hfil}\\
\end{tabular}}\\
\tableline
\end{tabular}
\begin{tabular}{r|r|r@{~$\pm$~}l|r|r|r@{~$\pm$~}l|r|r|r@{~$\pm$~}l|r|r}
\multicolumn{14}{c}{DQ-model}\\
$3000$&$292$&$19.2$&$3.7$&$151$&$-7$&$-19.8$&$4.2$&$80$&$68$&$0.6$&$5.6$&$58$&$-20$\\
$6000$&$916$&$5.6$&$1.6$&$186$&$74$&$-8.5$&$1.8$&$59$&$10$&$2.9$&$2.4$&$327$&$13$\\
$7000$&$1068$&$7.3$&$1.5$&$198$&$81$&$-5.9$&$1.7$&$77$&$4$&$-1.3$&$2.3$&$347$&$7$\\
$8000$&$1240$&$8.1$&$1.4$&$142$&$87$&$-6.2$&$1.5$&$102$&$-2$&$-1.9$&$2.1$&$12$&$2$\\
$8500$&$1322$&$6.8$&$1.3$&$68$&$85$&$-5.2$&$1.5$&$106$&$-4$&$-1.6$&$2.0$&$16$&$-3$\\
$9000$&$1379$&$7.0$&$1.2$&$53$&$79$&$-5.6$&$1.4$&$109$&$-6$&$-1.4$&$1.8$&$18$&$-9$\\
$9500$&$1417$&$6.9$&$1.2$&$41$&$78$&$-5.4$&$1.4$&$109$&$-4$&$-1.5$&$1.8$&$18$&$-11$\\
\hbox to \tmpl {\hfil $10000$}&$1459$&$6.3$&$1.2$&$55$&$75$&$-4.4$&$1.3$&$105$&$-10$&$-1.9$&$1.8$&$13$&$-11$\\
\multicolumn{14}{c}{DQO-model}\\
$3000$&$292$&$22.3$&$4.3$&$153$&$-9$&$-19.9$&$4.3$&$80$&$63$&$-2.4$&$6.1$&$59$&$-26$\\
$6000$&$916$&$5.1$&$1.9$&$156$&$71$&$-7.8$&$1.9$&$56$&$3$&$2.8$&$2.7$&$325$&$19$\\
$7000$&$1068$&$6.6$&$1.8$&$124$&$81$&$-5.4$&$1.7$&$75$&$-6$&$-1.2$&$2.5$&$346$&$7$\\
$8000$&$1240$&$7.8$&$1.6$&$141$&$86$&$-5.9$&$1.6$&$90$&$-3$&$-1.9$&$2.3$&$360$&$3$\\
$8500$&$1322$&$7.5$&$1.6$&$133$&$84$&$-4.9$&$1.5$&$92$&$-5$&$-2.6$&$2.2$&$2$&$4$\\
$9000$&$1379$&$8.2$&$1.6$&$101$&$84$&$-5.1$&$1.5$&$90$&$-6$&$-3.1$&$2.2$&$359$&$1$\\
$9500$&$1417$&$7.8$&$1.6$&$93$&$86$&$-4.8$&$1.5$&$99$&$-4$&$-3.0$&$2.2$&$9$&$-1$\\
$10000$&$1459$&$7.8$&$1.6$&$120$&$83$&$-5.3$&$1.5$&$89$&$-6$&$-2.5$&$2.2$&$359$&$3$\\
\tableline
\end{tabular}
\end{table*}

As a result, the variation of the subsample depth $R_{max}$ can lead to significant
deviation of the vector $\vec{D}$, especially when the supsample boundary
crosses a large attractor. At smaller $R_{max}$ values we observe only the increase of
radial velocity in the direction of the attractor. At larger $R_{max}$ we see
also the backfall. In the framework of DQO-model these effects can be taken into
account, but it is too much for the simple D-model to handle. On Fig.
\ref{fig:4} we show the variation of the apex position in dependence from the
sample depth. One can see that at small $R_{max}$ the apex crosses the galactic
plane and approaches the results of most authors depicted on Fig. \ref{fig:3}.
An additional argument in favour of the apex deviation being not much of a
problem is the following consideration. If in the framework of D-model with the
same $R_{max}$ for each galaxy we introduce statistical weights corresponding to
the variance of the radial velocity as a function of distance, the apex will
correspond to the position $l=327\deg;\,b=18\deg$, which is closer to the result
of \citet{ref:ParTug04}. It is obvious that in such an approach the input from
the nearest galaxies will dominate over the more distant galaxies.

Another possible explanation is that the velocity field is not well described by
a pure dipole, and the dipole one recovers is unstable with respect to the limiting
distance of the sample used. Furthermore, although the dipole is (in principle) a
well-defined property of the field, and its mean convergence with distance is
predictable within a given cosmological model, the problem here (as noted previously)
is that the multipole basis is not orthogonal for this sample, so that in practice
the dipole is not well-defined (in addition to varying stochastically with sample
volume). As \citep{ref:WFH09} have shown, this sample bias in the measurement of the
dipole is substantial, and has to be taken into account when comparing results from
different samples.

As we mentioned before, the usage of the more complex DQO-model allows to
describe such phenomena as the backfall. Besides, as it was shown by
\citet{ref:ParPar08}, the mean deviation of the calculated apex from its true
value, caused by measurement errors and the deviations from the Tully-Fisher
relationship, is less in the DQO-model in comparison to the D-model. On Fig.
\ref{fig:3} the relevant apex position for the DQO-model is denoted by a
diamond. We can see that it fairly corresponds to the results of other authors.

As it was mentioned above, we used the generalised Tully-Fisher relationship in
the form \ref{eqn:TF}. However, we also calculated the bulk motion parameters in
the D-model using the classical form \ref{eqn:TFalpha}. We obtained that the bulk
motion velocity is $274\,km\,s^{-1}$ and the apex coordinates are $l=317$, $b=36$
for $R_{max}=10000\,km\,s^{-1}$. For $R_{max}=8000\,km\,s^{-1}$ the corresponding
values are $254\,km\,s^{-1}$, $l=306$, $b=41$. As one can see from Table \ref{tbl:2}
and Fig. \ref{fig:3} these values fall in $1\sigma$ confidence areas.

Let us consider the quadrupole component. In Table \ref{tbl:3} we present the
corresponding coefficients. The traceless quadrupole tensor is
described by 5 coefficients:
\begin{equation}\label{eqn:Q}
\begin{array}{l}
V_r^{qua}=RQ_{ij}n_i n_j=R[q_1(n_1^2-n_3^2)+q_2(n_2^2\\
\phantom{V_r^{qua}=}{}-n_3^2)+q_3n_1n_2+q_4 n_1 n_3+q_5 n_2 n_3].
\end{array}
\end{equation}

\begin{figure*}[tb]
\includegraphics[width=\textwidth]{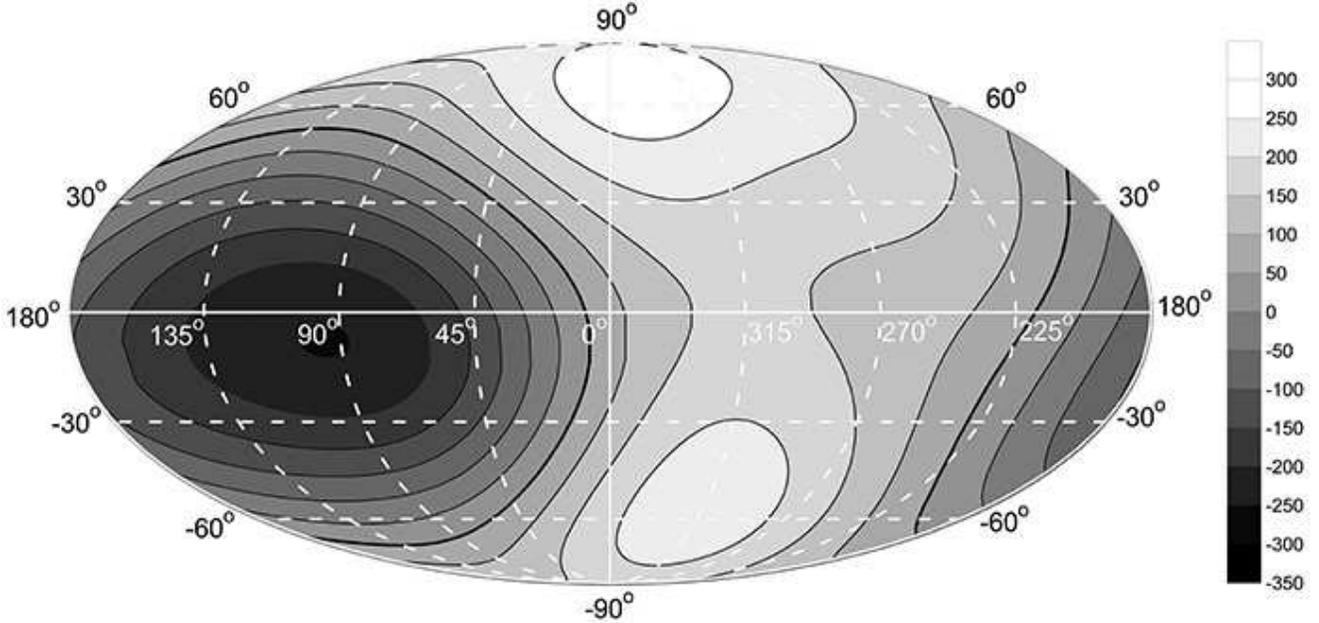}
\caption{Distribution of radial velocity of collective non-Hubble motion at $20h^{-1}\,Mpc$}
\label{fig:5a}
\end{figure*}

\begin{figure*}[tb]
\includegraphics[width=\textwidth]{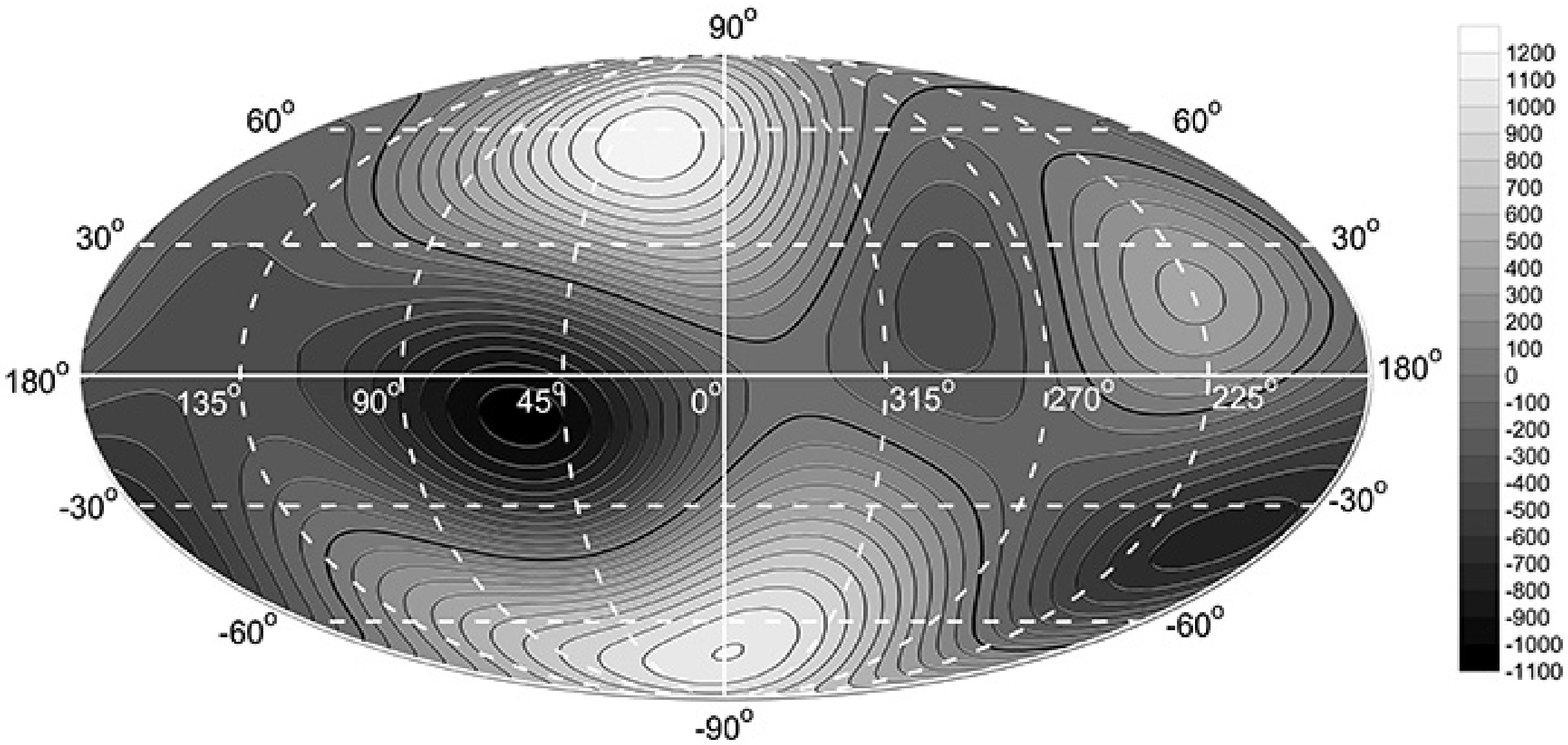}
\caption{Distribution of radial velocity of collective non-Hubble motion at $80h^{-1}\,Mpc$}
\label{fig:5}
\end{figure*}

\begin{table}[tb]
\settowidth{\tmpl}{Total}
\settowidth{\tmplj}{$\pm$}
\settowidth{\tmpla}{$-17.9$ $\pm$ $10.2$}
\settowidth{\tmplb}{$4.16$}
\settowidth{\tmplc}{$R_{max}=10000$}
\settowidth{\tmpld}{$4.37$}
\settowidth{\tmple}{$R_{max}=10000$}
\caption{Coefficients of the octopole component}
\label{tbl:5}
\begin{tabular}{c@{}r@{~$\pm$~}l|r|r@{~$\pm$~}l|r}
\tableline
\hbox to \tmpl {\hfil $i$\hfil}&
\multicolumn{6}{c@{}}{\begin{tabular}{|c|c|c|c}
\multicolumn{2}{|c|}{$R_{max}=8000$ km/s}&\multicolumn{2}{c}{$R_{max}=10000$ km/s}\\
\tableline
\hbox to \tmpla {\hfil $d_i\cdot 10^6$\hfil}&\hbox to \tmplb {\hfil $F_i$\hfil}&\hbox to \tmplc {\hfil $d_i\cdot 10^6$\hfil}&\hbox to \tmpld {\hfil $F_i$\hfil}\\
\end{tabular}}\\
\tableline
\end{tabular}
\begin{tabular}{r|r@{~$\pm$~}l|r|r@{~$\pm$~}l|r}
1&$7.2$&$3.9$&$3.5$&$-4.1$&$3.0$&$1.8$\\
2&$-2.1$&$5.2$&$0.2$&$-0.4$&$3.7$&$0.0$\\
3&$-3.7$&$5.4$&$0.5$&$-1.6$&$4.6$&$0.1$\\
4&$20.3$&$7.1$&$8.2$&$10.6$&$4.9$&$4.7$\\
5&$-7.8$&$7.4$&$1.1$&$4.3$&$6.1$&$0.5$\\
6&$24.0$&$7.6$&$9.8$&$23.0$&$5.5$&$17.5$\\
7&$-13.0$&$9.6$&$1.8$&$-12.5$&$8.0$&$2.4$\\
8&$10.9$&$7.8$&$2.0$&$15.8$&$6.1$&$6.8$\\
9&$-17.9$&$9.0$&$3.9$&$-6.2$&$6.9$&$0.8$\\
10&$33.8$&$10.2$&$11.1$&$25.9$&$7.6$&$11.7$\\
Total&\multicolumn{2}{c|}{}&$4.16$&\multicolumn{2}{c|}{\hbox to \tmple {}}&$4.37$\\
\tableline
\end{tabular}
\end{table}

\begin{figure}[tb]
\includegraphics[width=\columnwidth]{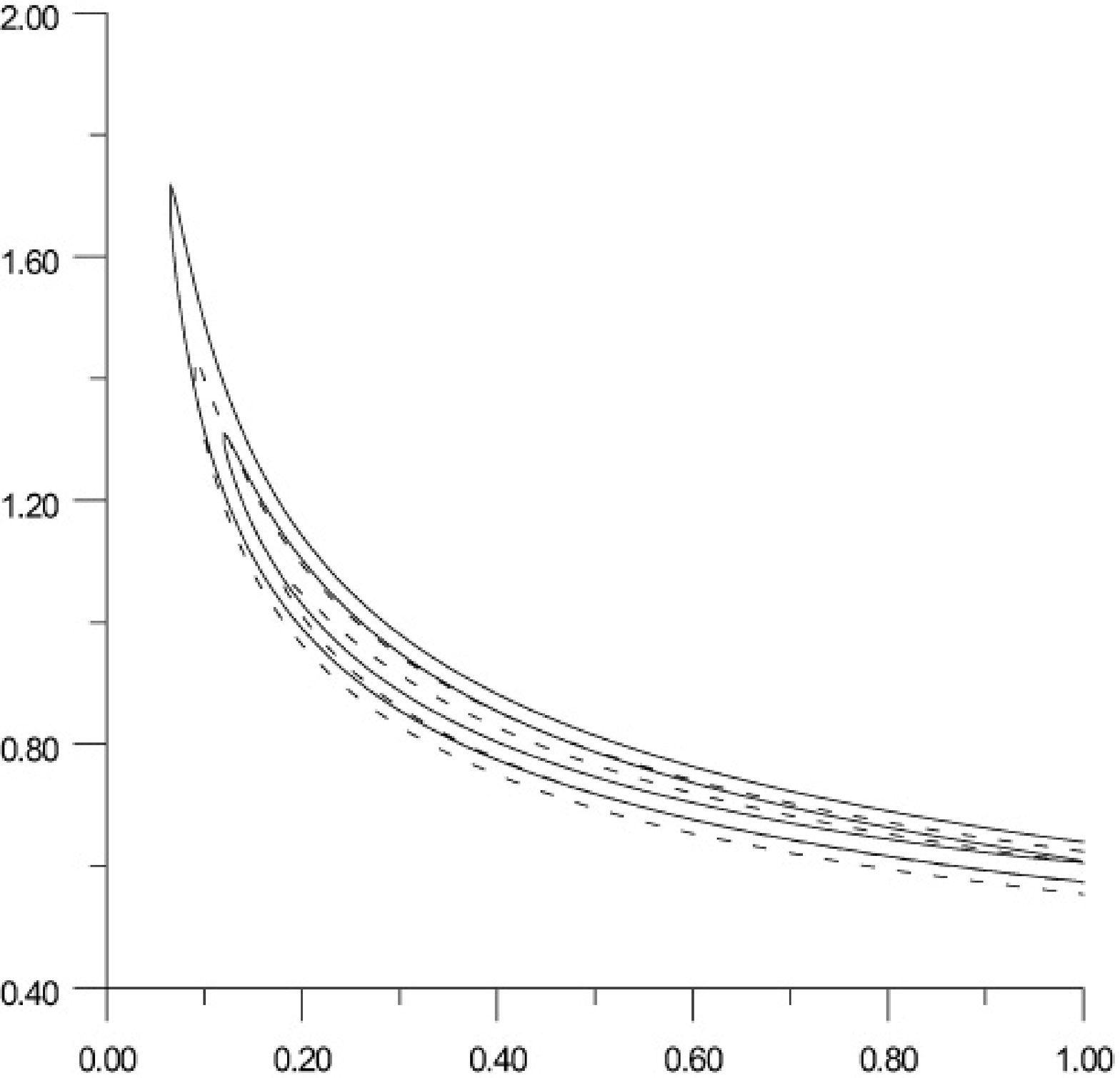}
\caption{$1\sigma$ and $2\sigma$ confidence levels of cosmological constants
$\Omega_m$ and $\sigma_8$ for the complex (solid) and simple (dashed) models}
\label{fig:6}
\end{figure}

In Table \ref{tbl:3} we also present the errors of these coefficients and their
significances according to F-test. The quadrupole coefficients $q_i$ are close
to the coefficients obtained by \citet{ref:ParTug04}. One can see that the
significance of all components except the first one is low. Nevertheless, the
significance of the first component is high enough for the total significance of
the whole quadrupole component to be sufficient. The $F$ values of the
quadrupole component are the following: $4.63$ for the DQ-model and
$R_{max}=10000\,km\,s^{-1}$, $4.85$ for the DQO-model and
$R_{max}=10000\,km\,s^{-1}$, $6.44$ for the DQ-model and
$R_{max}=8000\,km\,s^{-1}$, and $5.10$ for the DQO-model and
$R_{max}=8000\,km\,s^{-1}$. This should be compared to $3.35$, which corresponds
to the confidence level of $99.5$ per cent for 5 degrees of freedom. Thus the
quadrupole component is statistically significant according to F-test at $99.5$
per cent confidence level both in DQ and DQO-models.

What is the physical sense of the quadrupole component? As one can see from the paper
\citep{ref:Par01}, it can be naturally combined with the Hubble constant. As a result,
we obtain the effective `Hubble constant' depending on direction
\begin{equation}\label{eqn:HC}
H(l,b)=H(1+Q_{ik}n_i n_k).
\end{equation}
Naturally, this effective `Hubble constant' is caused by large-scale collective
motion on the sample scale. To estimate the value of its anisotropy we found
the eigenvalues and eigenvectors of tensor $Q$. The three eigenvectors are
orthogonal and the sum of three eigenvalues is equal to zero because $\tens{Q}$
is a traceless tensor.

In Table \ref{tbl:4} we present the maximum, minimum and
medium eigenvalues and the corresponding eigenvectors for DQ and DQO-models and
different $R_{max}$. To obtain the eigenvalues of this `Hubble constant' we
should add $100$ per cent to each of the listed values. For example, for the
subsample with $R_{max}=10000\,km\,s^{-1}$ in the framework of DQ-model the
maximal `Hubble constant' is equal to $106.3$ per cent of the actual Hubble
constant, and the minimal -- $95.6$ per cent. We can see that the anisotropy is
weak but statistically significant. The anisotropy is large only for very small
scales less than $60h^{-1}\,Mpc$. As one can see from the Table \ref{tbl:4}, the
maximum axis is dominant over the other two. However, the direction of the
minimum axis is more stable for both DQ- and DQO-model. In the DQO-model the
maximum axis is also stable at $R_{max}>6000\,km\,s^{-1}$. It lies almost in the
supergalactic plane. The minimum direction deviates from this plane by
approximately $30$ degrees of arc.

In the DQ-model the situation is different.
As one can see from the Table \ref{tbl:4}, the variation of the maximum
eigenvector is more evident. With the increase of $R_{max}$ the eigenvector
starts deviating from the supergalactic plane and at $R_{max}=10000\,km\,s^{-1}$
this deviation reaches approximately $20$ degrees of arc. This effect can be
explained by the fact that the Local Supercluster in this case constitutes the
smaller fraction of the sample.

We also calculated the statistical significance of the octopole component in the
DQO-model. The corresponding $F$ values are equal to $4.37$ for
$R_{max}=10000\,km\,s^{-1}$ and $4.16$ for $R_{max}=8000\,km\,s^{-1}$. These
values should be compared to the value $2.52$, which corresponds to the
confidence level of $99.5$ per cent for $10$ degrees of freedom. Thus, the
octopole component is also statistically significant according to F-test at
$99.5$ per cent confidence level. The same qualitative conclusion was achieved in the
paper \citep{ref:ParPar08}. We can also calculate the statistical significance of
$\vec{P}$ in eq. (\ref{eqn:P}) as in the paper \citep{ref:Par01,ref:ParTug04}. Its $F$ values are equal to $0.21$ for
$R_{max}=10000\,km\,s^{-1}$ and $2.50$ for $R_{max}=8000\,km\,s^{-1}$. This
should be compared to $3.78$, $2.60$, $2.14$ and $0.79$, which correspond
respectively to $99$, $95$, $90$ and $50$ per cent confidence levels for $3$
degrees of freedom. Thus, for $R_{max}=10000\,km\,s^{-1}$ the octopole trace is
insignificant at $50$ per cent confidence level, and for
$R_{max}=8000\,km\,s^{-1}$ it is significant at $90$ per cent confidence level,
but this is insufficient to claim that this value is significant for all
subsamples.

In Table \ref{tbl:5} we present the coefficients of the octopole component,
their errors and significances according to F-test. The octopole tensor
including the trace $\vec{P}$ is described by $10$ coefficients:
\begin{equation}\label{eqn:O}
\begin{array}{l}
V_r^{oct}=R^2\hat{O}_{ijk}n_in_jn_k=R^2(d_1n_1^3+d_2n_2^3\\
\phantom{V_r^{oct}=}{}+d_3n_3^3+d_4n_1n_2^2+d_5n_1n_3^2+d_6n_2n_1^2\\
\phantom{V_r^{oct}=}{}+d_7n_2n_3^2+d_8n_3n_1^2+d_9n_3n_2^2\\
\phantom{V_r^{oct}=}{}+d_{10}n_1n_2n_3).
\end{array}
\end{equation}

The coefficients of the octopole component differ from those given by
\citet{ref:ParTug04} much stronger than the quadrupole ones. Nevertheless, their
$1\sigma$ confidence areas overlap.

The knowledge of the coefficients of dipole, quadrupole and octopole components
allows us to calculate the distribution of the radial velocity of the collective motion
of galaxies. On Fig. \ref{fig:5a} we present such a distribution for the
DQO-model with $R_{max}=10000\,km\,s^{-1}$ and $R=2000\,km\,s^{-1}$, and on Fig. \ref{fig:5}
-- for $R_{max}=10000\,km\,s^{-1}$ and $R=8000\,km\,s^{-1}$. Note that
the amplitudes of the dipole, quadrupole and octopole components are close to
each other at $R\sim 6000\,km\,s^{-1}$. Thus, for larger distance like in Fig.
\ref{fig:5} the octopole component prevails in the velocity field. This figure
is characterised by a deep minimum of radial velocity in the direction opposite
to the Great Attractor and Shapley concentration. In the direction to these
superclusters there is also a minimum caused by a backfall to the Great
Attractor. For smaller distances $R$ like in Fig. \ref{fig:5a} this minimum
becomes a maximum, but the
minimum in the opposite direction is deeper than this maximum. It is clear that
this picture is a simplified representation of the velocity field, but still it
is much more complex than the simple bulk motion given by the D-model.

\section{Estimation of some cosmological parameters and their combinations}\label{s:Cosmology}

As a result, we obtain not only the parameters of the multipole models of the collective
motion, but also an estimation of the distance to galaxies using the generalised
Tully-Fisher relationship. This allows us to compile a list of peculiar
velocities for $1623$ galaxies. It can be used to estimate the cosmological
parameters $\Omega_m$ and $\sigma_8$. For the previous version of the sample it
was done in the paper \citep{ref:cosm}. We use a method similar to that used by
\citet{ref:Feldman} but with some changes. We calculate a 2-point correlation
function for peculiar velocities, which is approximated by a formula obtained by
\citet{ref:Juszk}. The formula has a nonlinear dependency on $\Omega_m$ and
$\sigma_8$, thus these parameters can be estimated by the maximum likelihood
method using the correlation function for RFGC galaxies. This yields an
estimation of cosmological parameters $\Omega_m$ and $\sigma_8$ depicted on Fig.
\ref{fig:6}. There are 2 versions of the approximation formula: a simple one
and a complex one. For the complex formula we obtain the values
$\Omega_m=0.26$ and $\sigma_8=0.96$. This point is surrounded by $1\sigma$ and
$2\sigma$ confidence areas. They have the banana-like shapes -- very long and
narrow curved strips. Judging from the boundaries of these areas, the
errors of the cosmological parameters are very large. For this reason, the good
agreement in the value of $\Omega_m$ between our estimation and other
estimations can be just incidential. The simple model yields similar bananas.

\begin{figure}[tb]
\includegraphics[width=\columnwidth]{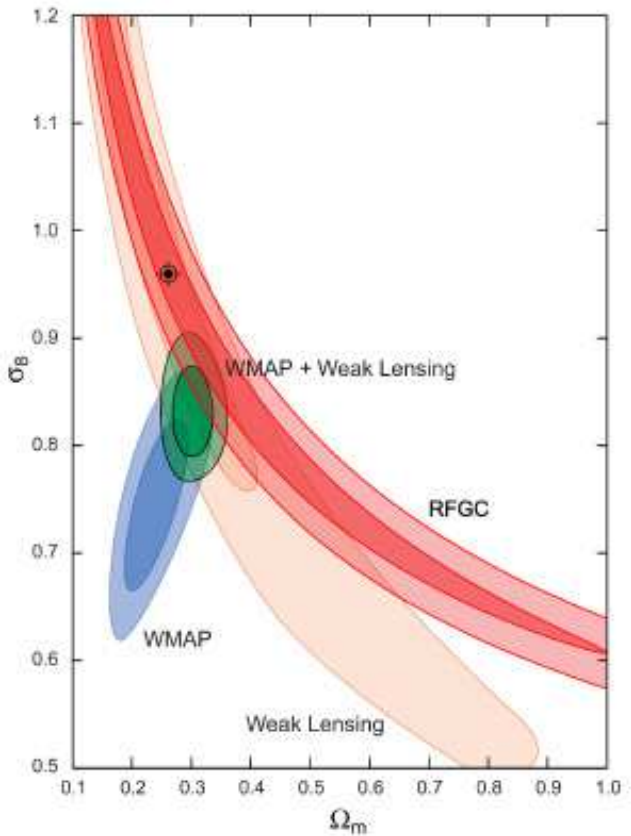}
\caption{An estimation of cosmological constants $\Omega_m$ and $\sigma_8$ superimposed on 3-year WMAP results}
\label{fig:7}
\end{figure}

\begin{figure}[tb]
\includegraphics[width=\columnwidth]{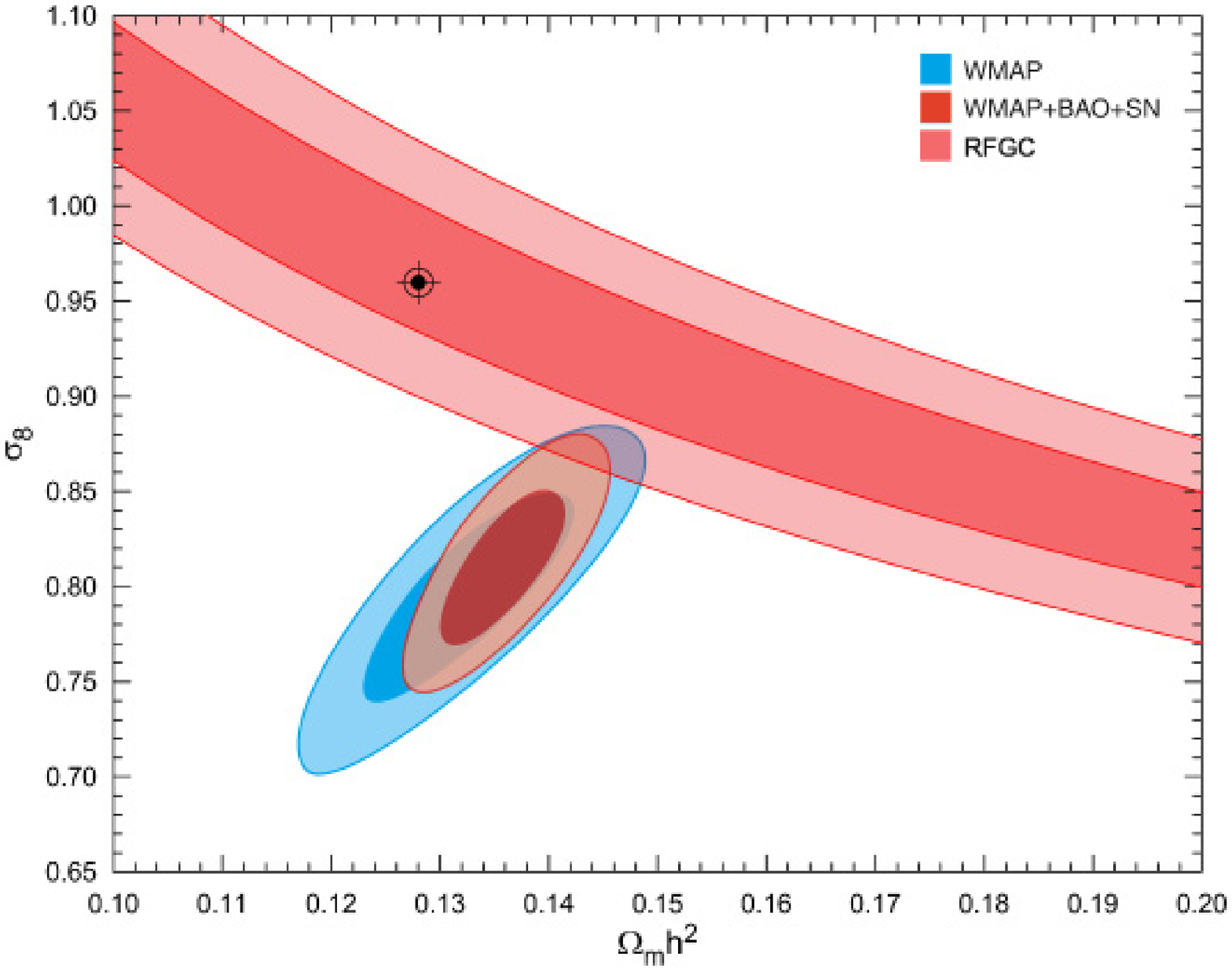}
\caption{An estimation of cosmological constants $\Omega_m$ and $\sigma_8$ superimposed on 5-year WMAP results}
\label{fig:8}
\end{figure}

Let us compare the obtained confidence areas with more accurate estimations. Let
us start from the 3-year WMAP results. On Fig. \ref{fig:7} we reproduce Fig. 7
from the paper \citep{ref:WMAP3} with an overlaid plot of $1\sigma$ and $2\sigma$
confidence areas for the complex model. The blue colour depicts WMAP data,
the yellow colour depicts
data obtained from weak gravitational lensing, the green colour depicts
constraints imposed by both WMAP and lensing, and the red colour depicts our
constraints. We can see that our results are similar to the weak lensing data
much more than to WMAP data. In any case, these results are obtained with
totally different methods, and their good correspondence is very promising.

The results of comparison with 5-year WMAP data are presented on Fig.
\ref{fig:8}. It is a reproduction of Fig. 19 from the paper \citep{ref:WMAP5} with an overlaid
plot of $1\sigma$ and $2\sigma$ areas for the complex model. It contains
constraints set by WMAP, mutual constraints set by WMAP, BAO and supernovae, and
our constraints.

\begin{figure}[tb]
\includegraphics[width=\columnwidth]{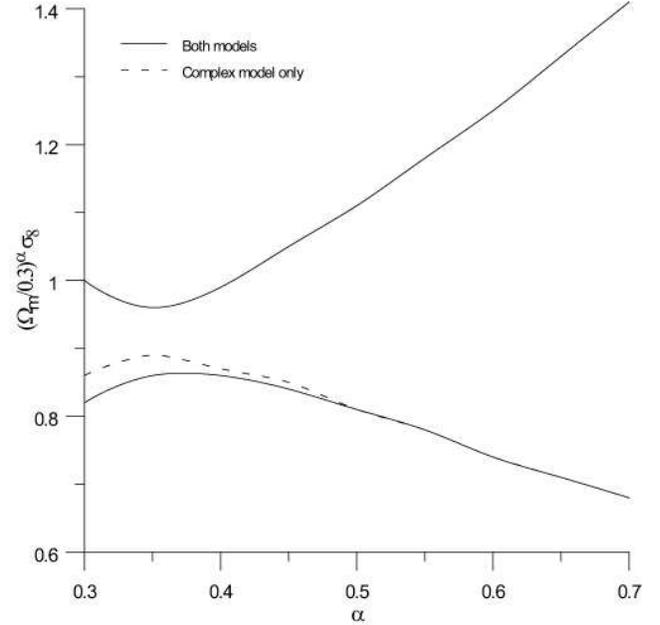}
\caption{Dependence of $(\frac{\Omega_m}{0.3})^\alpha \sigma_8$ on $\alpha$.
The solid line depicts constraint set by both models,
the dashed line depicts constraint set only by the complex model only}
\label{fig:9}
\end{figure}

The main drawback of our estimation is its graphical form. In
many cases it is preferable to deal with numerical constraints. To obtain such
constraints we will make use of the long and narrow shape of our confidence
areas. If we consider a combination of cosmological parameters of the form
$(\Omega_m/0.3)^\alpha \sigma_8$ and $\alpha$ in the range from $0.3$
to $0.7$ we will obtain an estimation of the combination with better accuracy
then either of the two cosmological parameters. The corresponding estimations
are given on Fig. \ref{fig:9}.

\begin{figure}[tb]
\includegraphics[width=\columnwidth]{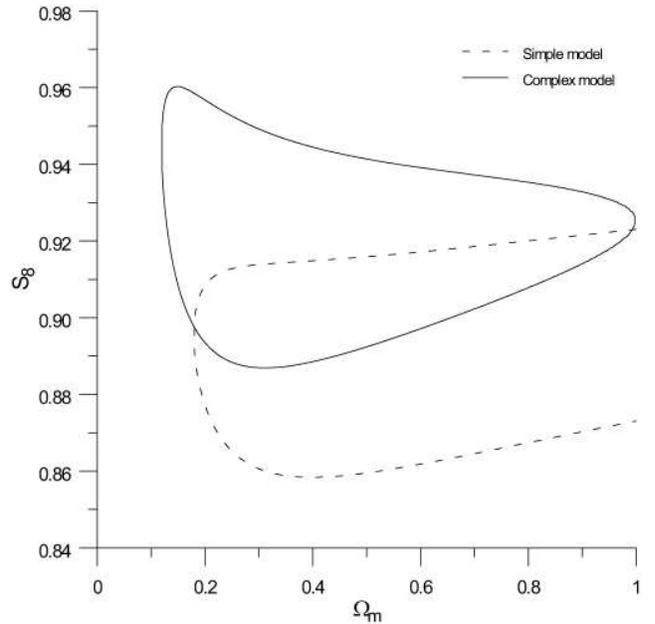}
\caption{$1\sigma$ confidence areas of $S_8$ vs. $\Omega_m$ dependence.
The solid line corresponds to the complex model,
the dashed line corresponds to the simple model}
\label{fig:10}
\end{figure}

For the previous version of the sample the estimation of such
a combination was made in the paper \citep{ref:Par08}. Ibid there are many
estimations of other authors for different values of $\alpha$. Of all the values of
$\alpha$ the smallest error was reached at $\alpha=0.35$. On Fig. \ref{fig:10}
one can see the boundaries of $1\sigma$ confidence areas for the complex
and simple models in the $(\Omega_m/0.3)^{0.35} \sigma_8$ against
$\Omega_m$ coordinates. Both of them fit the constraint
$(\Omega_m/0.3)^{0.35} \sigma_8=0.91\pm 0.05$. This result is almost the same as
for the previous version of the sample ($0.93\pm 0.05$). The importance of this
$\alpha$ value is caused by the fact that the value
$S_8=(\Omega_m/0.3)^{0.35} \sigma_8$ was used in the paper \citep{ref:Evrard}. As
the author notes, the low estimation $S_8=0.69$ based upon WMAP and SDSS
galaxies clusterisation leads to a number of problems. It leads to a
contradiction with results of clusters formation modelling. The high estimation
$S_8\sim0.8-0.9$ lifts these problems. The estimation of the share of hot gas
following from it corresponds to modern data obtained from the
Sunyaev-Zel'dovich effect. Our results support the high estimation. Note
that the paper \citep{ref:Evrard} contains many estimations obtained by different
authors with different methods. Among them there are both high estimations,
close to ours, and low estimations.

\section{Conclusion}\label{s:Conclusion}

We prepared a new increased and largely revised sample of RFGC galaxies with
data about redshifts and HI linewidths. It allowed us to improve the estimations
of parameters of the radial velocity field of large-scale collective motions of
galaxies using 3 models of its multipole structure. In comparison with the previous
versions the main features remained intact, but separate details, for example
the apex of the dipole component of the bulk motion in the D-model, significantly changed. As
before, the statistical significance according to F-test of both the quadrupole and the octopole
components is well above $99.5$ per cent. The exact parameters are given in the
article for subsamples with maximum distances $100h^{-1}\,Mpc$ and $80h^{-1}\,Mpc$.

The obtained velocity of the bulk motion is in agreement with the $\Lambda$CDM model
expectation of $\sim 200\,km\,s^{-1}$. The value of the bulk motion
velocity attracted additional attention after the recent paper of \citet{ref:WFH09} who
obtained this value as $407 \pm 81\,km\,s^{-1}$ at the same scale $100h^{-1}\,Mpc$ as
in present article. In this connection some authors immediately started speculating
about the challenge to the $\Lambda$CDM model and the necessity for inclusion of
non-gravitational forces \citep{ref:AWW09}.

For $1623$ galaxies we compiled a list of peculiar velocities which will be
published in the nearest future. We also intend to use it to determine the
distribution of the matter density (including dark matter) at the scales
$75h^{-1}\,Mpc$ using POTENT method. This list was also used to estimate the
cosmological parameters $\Omega_m$ and $\sigma_8$. The obtained constraint is
given in graphical form on Fig. \ref{fig:6}. The best numerical constraint is
given by a combination $S_8=(\Omega_m/0.3)^{0.35} \sigma_8=0.91\pm 0.05$.

\acknowledgments

We acknowledge the usage of the HyperLeda data\-base (http://leda.univ-lyon1.fr).

This research has made use of the NASA/IPAC Extragalactic Database (NED) which
is operated by the Jet Propulsion Laboratory, California Institute of
Technology, under contract with the National Aeronautics and Space
Administration.

This article is published in the framework of Programme ``Cosmomicrophysics'' of
the National Aca\-demy of Sciences of Ukraine.

\end{document}